\documentclass[12pt]{iopart}

\usepackage{graphicx}
\usepackage{cite}

\usepackage{iopams}
\usepackage[usenames,dvipsnames]{color}
\usepackage{hyperref}
\hypersetup{
colorlinks=true,
citecolor=blue,
filecolor=blue,
linkcolor=DarkOrchid,
urlcolor=Blue
}

\usepackage[small,bf]{caption}

\usepackage[scaled]{helvet} 
\usepackage{courier} 
\usepackage[euler-digits]{eulervm}

\begin{document}

\title{Passive, free-space laser gyroscope}
\author{W.~Z.~Korth, A.~Heptonstall, E.~D.~Hall, K.~Arai, E.~K.~Gustafson and R.~X.~Adhikari}
\address{LIGO Laboratory, California Institute of Technology, Pasadena, CA  91125}
\ead{korth@caltech.edu}

\begin{abstract}
Laser gyroscopes making use of the Sagnac effect have been used as highly accurate
rotation sensors for many years. First used in aerospace and defense 
applications, these devices have more recently been used 
for precision seismology and in other research settings. In particular, mid-sized 
($\sim1$ m-scale) laser gyros have been under development as tilt sensors to augment 
the adaptive active seismic isolation systems in terrestrial interferometric 
gravitational wave detectors. The most prevalent design is the 
``active'' gyroscope, in which the optical ring cavity used to measure the Sagnac 
degeneracy breaking is itself a laser resonator. In this article, we describe another 
topology: a ``passive'' gyroscope, in which the sensing cavity is not itself a laser 
but is instead tracked using external laser beams. While subject to its own 
limitations, this design is free from the deleterious lock-in effects observed 
in active systems, and has the advantage that it can be constructed using 
commercially available components. We demonstrate that our device achieves 
comparable sensitivity to those of similarly sized active laser gyroscopes.
\end{abstract}

\clearpage
\section{Introduction}
\label{s:intro}

\subsection{Laser Gyroscopes}
The first use of a ring laser cavity to detect rotational motion was demonstrated by Macek and 
Davis\,\cite{Macek:1963fk} in 1963, and their design remains essentially unchanged in most current 
implimentations. The operating principle for all optical gyroscopes is the {\it Sagnac effect}: in a 
ring geometry, if the system is rotating in the optical plane, the roundtrip optical path lengths 
traversed by two counter-propagating beams are unequal. 

This earliest design was itself an improvement of a non-resonant, phase-sensitive interferometer introduced 
by Sagnac himself\,\cite{Sagnac:1913uq}. In that instrument, an interferometric fringe shift was produced at 
the output proportional to the rotation rate, and its sensitivity was therefore limited by the achievable 
fringe resolution\footnote{Another type of interferometric optical gyroscope, the {\it fiber optic gyroscope (FOG)},
 is not discussed here. It is similar to Sagnac's original interferometer, but with the free-space system replaced 
by many windings of a fiber, increasing sensitivity. For an excellent contemporary review of all optical gyroscope 
technologies, see\,\cite{Chow:1985fr}.}. Conversion of the ring into a laser cavity created a bidirectional resonator, 
wherein the supported modes in each of the two directions---their frequencies being dependent on the respective 
roundtrip phases---are non-degenerate in the presence of rotation. This allowed Macek and Davis to use far more 
sensitive heterodyne techniques to measure the frequency splitting caused by rotation (at the time of their work, 
the achievable resolution was one part in $10^{12}$, a significant improvement over the interferometric fringe readout). 
As reported in their original paper, the Sagnac-induced frequency shift is given 
by\footnote{This relationship can easily be proven for a circular path, and holds true for an arbitrary geometry.}:
\begin{equation}
\Delta \nu = \frac{4}{\lambda S} \vec A \cdot \vec \omega,
\label{eq:rads2hz}
\end{equation}
where $\vec A$ is the vector of area enclosed by the cavity, $S$ is the cavity perimeter, $\lambda$ is the 
laser wavelength, $\vec \omega$ is the angular velocity, and $\Delta \nu$ is the optical frequency splitting.

The concept of an externally illuminated (``passive'') laser gyroscope was first presented by Ezekiel and 
Balsamo\,\cite{Ezekiel:1977kx} in 1977. Previously, a major issue with the common active design had been 
discovered: at small rotation rates, backscatter-induced crosstalk effects caused the counter-propagating modes to lock to one another 
in frequency, leading to a null output\footnote{Upon the later construction of large enough units, it was found 
that the DC Sagnac shift afforded by the earth's rotation was enough to prevent the ``lock-in'' effect. A 
calculation for the required gyroscope size in order to avoid lock-in, as a function of backscatter coefficient, 
can be found in\,\cite{Belfi:2010pb}.}. It was believed that this effect was caused by the presence of the gain 
medium within the gyroscope cavity\,\cite{Ezekiel:1977kx,Aronowitz:1971vn}. Ezekiel and Balsamo sought therefore 
to circumvent this effect by locking an external laser to a passive optical ring cavity. In their setup, 
acousto-optic modulators (AOMs) were used to shift the laser frequency up macroscopically in common mode for the 
two counter-propagating beams. A primary loop locked the cavity length to one upshifted beam, and a secondary 
loop adjusted the frequency of the other beam's AOM to lock it to the counter-propagating mode. Ultimately, it 
was found that even passive designs exhibit this ``lock-in'' effect\,\cite{Zarinetchi:1986mz, Zarinetchi:1992ue}, 
which was determined to be the result of back-scattering from one beam to the other.

In this article, we report on a variant of the passive design by Ezekiel and Balsamo. The salient departure from 
that design is the operation of the two counter-propagating beams on adjacent axial modes of the ring cavity, 
such that the two fields are separated in frequency by one cavity free spectral range (100 MHz, in our case). 
This macroscopic frequency separation reduces the intracavity crosstalk, allowing for enhanced
high-sensitivity, linear operation of the gyroscope down to zero frequency. 

\subsection{Current laser gyroscope sensitivities}
\label{sec:other_gyros}
Theoretically, due to the dimensional dependence in Eq.\,\ref{eq:rads2hz}, larger-area 
gyroscopes are inherently more sensitive than smaller ones. In applications where size is not a 
great concern, such as in geophysical experiments, the paradigm has been to make instruments as large 
as is practical. Several large geophysical
gyroscopes\,\cite{Stedman:1997gf, Schreiber:2003ul, Schreiber:2004pd, Rowe:1999lq, 
Dunn:2002rr,Hurst:2009wd} 
have in recent years demonstrated a resolution of one part in $10^8$ of the earth's rotation rate over 
several-hour integration times. The largest of these gyroscopes is ``UG-2''\,\cite{Hurst:2009wd}, a 
rectangle with $39.7\mbox{ m} \times 21$\,m sides.

On the other end of the spectrum, more compact designs have been used in aerospace for decades as 
an important component of inertial guidance systems. Currently available 
models\,\cite{Honeywell:qe,L3:ai}, typically 10--20\,cm on a side, exhibit best noise levels of 
$\sim10^{-6} - 10^{-5}$ (rad/s)/$\sqrt{\mbox{Hz}}$ and DC stability of 0.001--0.01\,$^{\rm o}$/hr.

In recent years, the use of laser gyroscopes has been investigated as a potential supplement to the active 
feedforward seismic isolation systems of the second-generation interferometric gravitational wave (GW) 
detectors Advanced LIGO\,\cite{Lantz:2009hc,DeRosa:2012ij} and Advanced VIRGO\,\cite{VIRGO:2009fv}. 
In those systems, a network of seismometers is used to sense ground motion around the GW 
interferometers' test masses, and their signals are used---via Weiner-filter based feedforward noise
cancellation---to subtract the ground motion with force actuators. It is well known that seismometers 
exhibit a parasitic sensitivity to ground tilt at low frequencies\,\cite{Lantz:2009hc}, and the concept is 
therefore to use rotation sensors in parallel to remove the spurious tilt-induced component of the 
ground motion signal. In this application, cost and space constraints and the desire for localized tilt 
information near multiple interferometer components dictate that the size of the used gyroscope 
be on the order of a meter.

In the case of Advanced VIRGO, a prototype laser gyroscope\,\cite{Belfi:2010pb,Belfi:2012fu}, ``G-Pisa'', 
has been constructed. Using the conventional active design with a modest size of 1.4\,m on a side, 
the G-Pisa sensor has been operating for some time at a sensitivity level of 
$\sim10^{-8} - 10^{-9}$ (rad/s)/$\sqrt{\mbox{Hz}}$. Based on a theoretical model for the 
active system, the Pisa group shows that this current noise floor is dominated by backscatter effects.

In this article, we will give a detailed description of a prototype passive, free-space laser 
gyroscope of 75-cm side length constructed with the aim of serving as a tilt sensor in the 
Advanced LIGO seismic isolation scheme. While not meeting these stringent requirements, we achieved a sensitivity of $10^{-8}$ (rad/s)$/\sqrt{\mbox{Hz}}$ above 500\,mHz.


\subsection{Rotation Sensing}
\label{sec:RotSens}
\begin{figure}[h]
\centering
\includegraphics[width=\columnwidth]{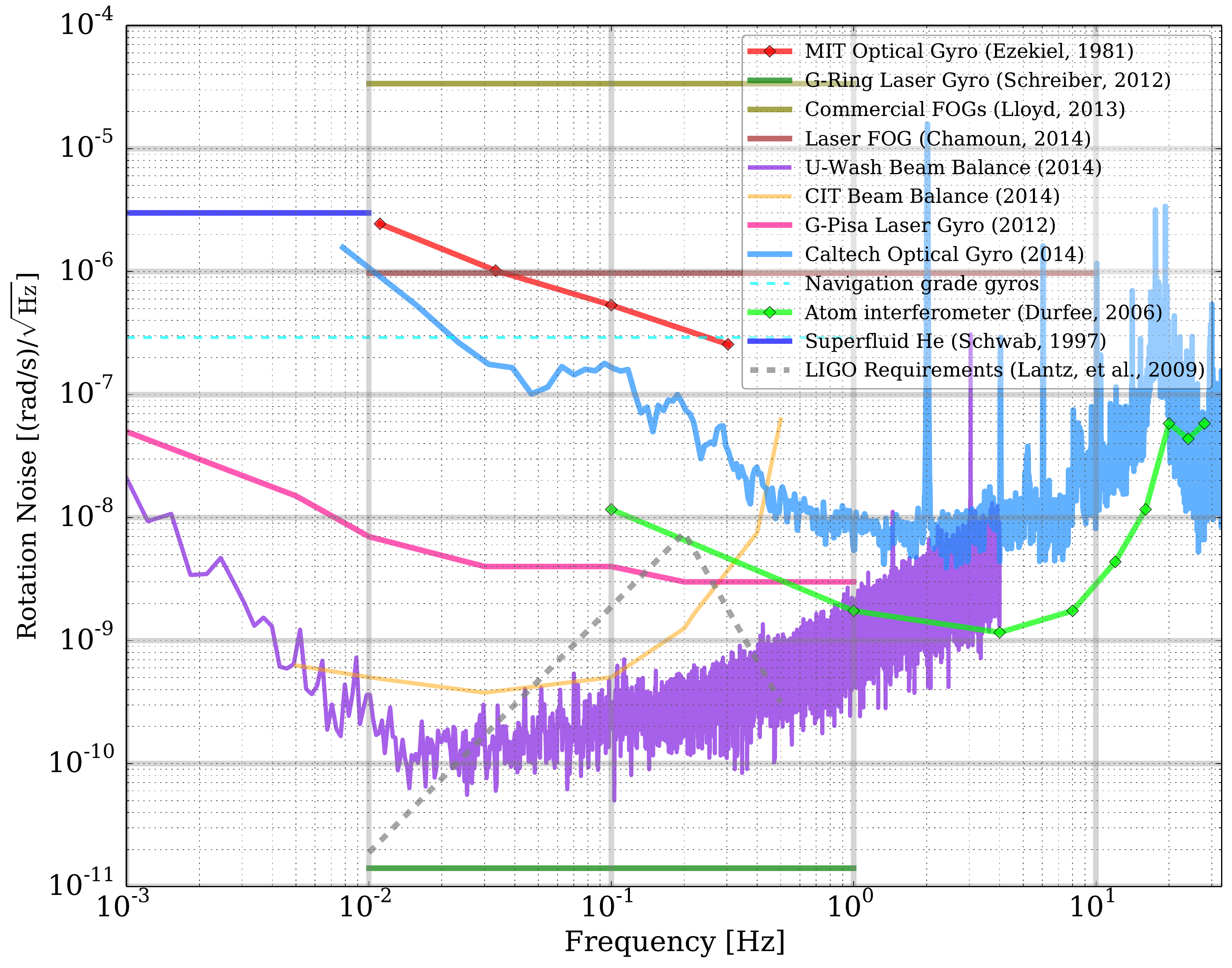}
\caption{Shown here are requirements for a number of different
	rotation sensing application as well as several types of
	rotation sensor:
	the MIT passive ring resonator~\cite{Sanders:1981gc},
	the U. Wash. balance beam~\cite{washu:2014}, 
	the Caltech balance beam~\cite{Vladimir:Tilt14},
	the G-Ring laser gyro~\cite{hadziioannou2012examining, Schreiber:2013j},
	the Laser FOG~\cite{chamoun2014low},
	the G-Pisa laser gyro~\cite{Belfi:2011er},
	the Stanford atom interferometer~\cite{Durfee:2006fk},
	and the UC-Berkeley superfluid He sensor~\cite{Schwab:1997uq}.}
	\label{fig:comp_plot}
\end{figure}

In addition to the active and passive, free-space laser gyros described above, a number of other
rotation sensor technologies exist. In Figure~\ref{fig:comp_plot}, we have compiled the
angular sensitivity of a number of these in order to place the requirements for the
gravitational-wave detectors in the proper context.

One of the earliest efforts to subtract tilt from the suspension point of a pendulum suspension
for interferometers was proposed by Robertson, et al.\,\cite{robertson1982passive} and utilized
a rotational reference arm~\cite{giazotto1986one}. Recently a few groups have demonstrated
low noise tilt sensing using balance beam sensors~\cite{washu:2014, Vladimir:Tilt14}. These
sensors are quite close to the sensitivity needed to reduce tilt at the upper stages of the LIGO
seismic isolation system, but do not address the issue of tilts generated within the isolation
system. For that, one would need to place a rotation sensor at the pendulum's suspension point
or to use a tilt-free sensor~\cite{citeulike:13347071} for the intertial isolation.

The fiber-optic ring gyro (FOG)~\cite{merlo2000fiber} has seen rapid development in the last few decades and is much more sensitive than the MEMS gyros used in mobile devices, but is not yet competitive with the best active laser gyros. Further improvements in reducing coherent backscatter and polarization modulation may yield an order of magnitude improvement, but this would still fall far short of modest free space laser gyros as well as the tilt sensing requirements of gravitational wave detectors.

Atom interferometers~\cite{1997PhRvLGust, gustavson2000precision, Durfee:2006fk} have also been used recently to make sensitive rotation measurements. In these systems, a laser is first used to prepare a beam of atoms into a known state, and then subsequently to interrogate the atomic state after some predetermined period of time. The rotation of the system can then be inferred through its well-understood coupling to the probability amplitudes of the output states. 
Another novel technique involves measuring the quantized flow of superfluid helium~\cite{Schwab:1997uq, PhysRevLett.78.3602}.

The sensitivity and noise of the various types of rotation sensors are 
compared in Fig.~\ref{fig:comp_plot}.


\clearpage
\section{The Passive, Free-Space Laser Gyroscope}

\subsection{Overview}
\label{sec:overview}

\begin{figure}[h]
\centering
\includegraphics[width=\columnwidth]{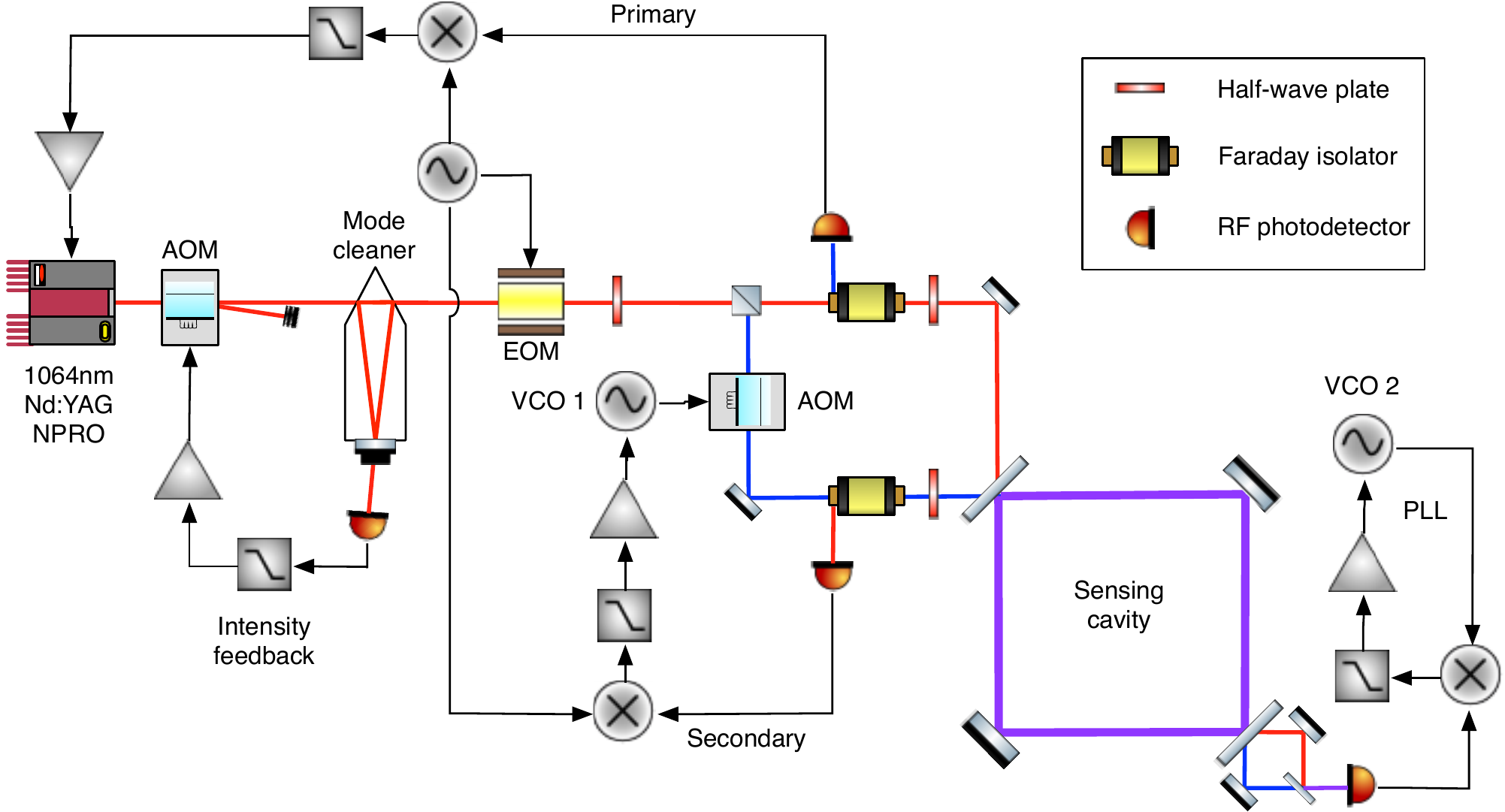}
\caption{Simplified diagram of the laser gyroscope. After mode cleaning and intensity stabilization, the main laser is locked to the counterclockwise mode of the sensing cavity. A pickoff of the beam is upshifted macroscopically by 100 MHz---roughly the FSR of the cavity---and feedback is applied to the AOM to lock this upshifted beam to the clockwise mode of the cavity. The rotation signal is encoded in both the control signal to the AOM as well as the beat between the main and secondary beams in transmission of the cavity.}
\label{fig:exp_diagram}
\end{figure}

A schematic diagram of the experiment can be found in Fig.\,\ref{fig:exp_diagram}. Light from a commercial, 
1064-nm Nd:YAG non-planar ring oscillator (NPRO) is locked to the counterclockwise mode of a 
square optical cavity via feedback to the laser frequency using the Pound-Drever-Hall (PDH) 
frontal-modulation heterodyne locking technique\,\cite{Drever:1983fk}. A pickoff of the input laser 
is upshifted using an acousto-optic modulator (AOM) by 100\,MHz, the free spectral range (FSR) of the 
cavity. By using a separate PDH loop to feed back to the AOM frequency, this upshifted beam is locked to 
the clockwise mode of the same ring cavity. In this configuration, using Eq.\,\ref{eq:rads2hz}, the 
frequency difference between these two beams is
\begin{equation} \label{eq:delta_nu}
\nu_{\rm cw} - \nu_{\rm ccw} = \frac{c}{S} + \frac{4}{\lambda S} \vec A \cdot \vec \omega,
\end{equation}
where $c$ is the speed of light and the area $\vec A$ is defined as vertically oriented.

The rotation signal can be read out in either of two ways. Most simply, the signal is encoded directly into the AOM 
actuation signal. Alternatively, the frequency difference can be measured by recombining the transmitted 
beams and measuring the beat in the detected photocurrent using a phase-lock loop (PLL). In either case, 
the output signal is directly sensitive to phase noise in one of two RF voltage-controlled oscillators (VCOs): 
in the former case, it is the VCO driving the AOM; in the latter, it is the VCO in the PLL. In each case, the 
small rotation-induced fluctuations are impressed on a 100\,MHz carrier. Using Eq.\,\ref{eq:rads2hz}, the target 
sensitivity of $\sim10^{-9}$ (rad/s)/$\sqrt{\mbox{Hz}}$ corresponds to an optical frequency shift of approximately 
1~mHz/$\sqrt{\mbox{Hz}}$, giving a required relative stability of $10^{-11}\, 1/\sqrt{\mbox{Hz}}$. Given the 
stability of currently available RF VCOs, this requirement puts a considerable constraint on the near-term 
improvement of the design we consider here\footnote{It should be noted that in the optical beat and PLL 
readout case, one is free to use a wider variety of frequency discriminators, some of which may in principle 
be inherently more stable.}.

\subsection{Detailed Experimental Design}

\begin{figure}[h]
\centering
\includegraphics[width=\columnwidth]{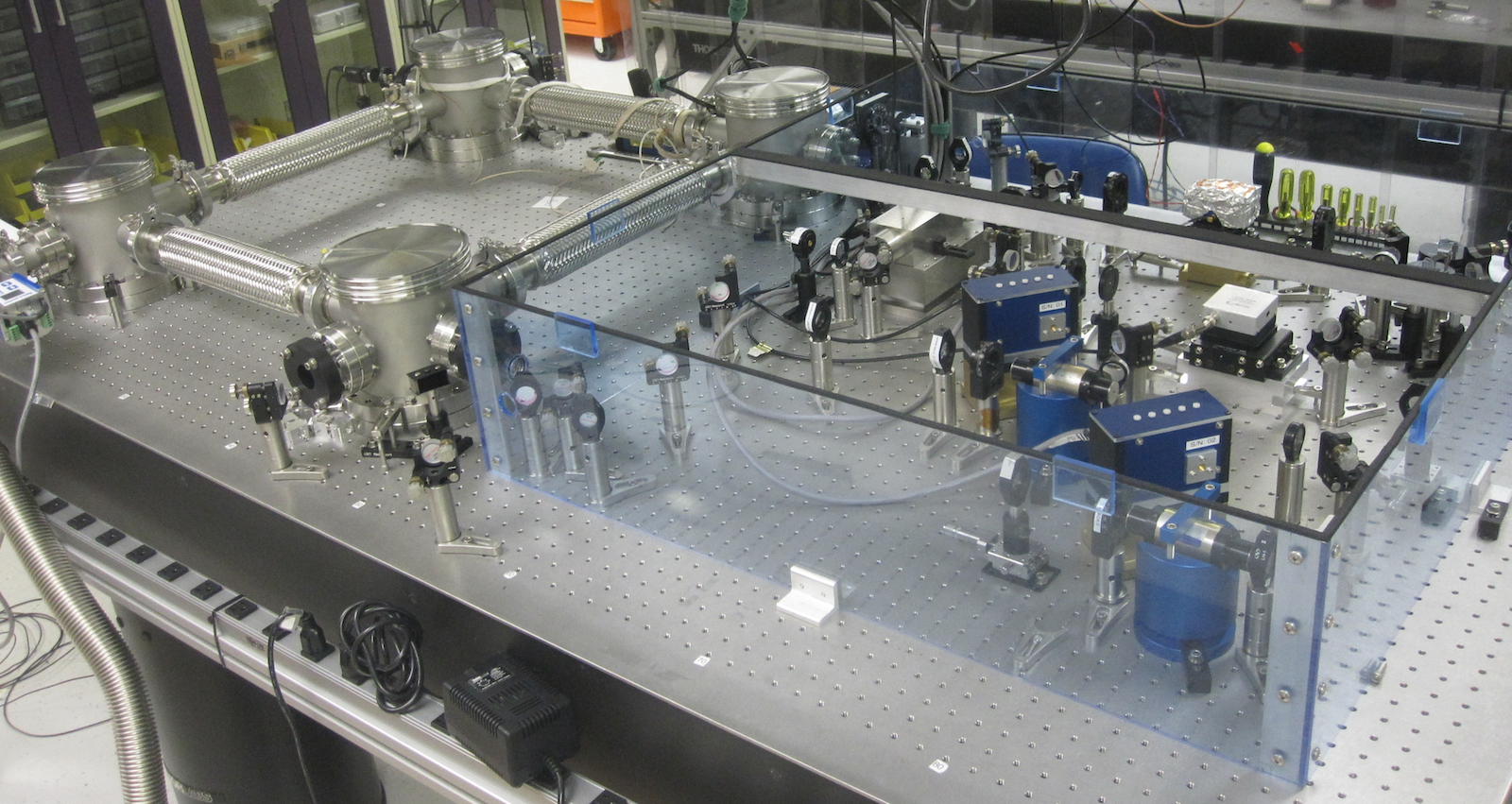}
\caption{Photo of the experiment}
\label{fig:photo}
\end{figure}

\subsubsection{Optical Cavity}
The core of the gyroscope is a square, free-space optical cavity of 75-cm side length. The cavity is critically 
coupled, with the input- and output-coupling mirrors each being 2-in diameter, flat, high-reflectivity (HR) optics positioned 
at opposing corners of the square. The cavity is geometrically stabilized using a single 3-m radius of curvature, 1-in 
HR mirror in another corner. The final optic is a flat, HR turning mirror. The coupling mirrors' power transmission of 
200\,ppm, along with the other mirrors' transmission and aggregate scatter and absorption losses, result in a finesse 
of approximately 12000. 

The cavity is enclosed in a custom vacuum system composed of steel corner chambers connected by flexible metal braid 
tubes using KF flanges. The optical signals are injected and extracted from the vacuum envelope through optical-quality 
wedged, anti-reflection (AR) coated windows. The purpose of the vacuum system is to remove optical path fluctuations 
induced by index of refraction variations of the air. As such, high vacuum is not required for low-noise 
operation; in practice, the chamber is evacuated to the mTorr level, sealed, and the pump removed to avoid 
excess vibration.

\subsubsection{Beam preparation and cavity injection/extraction}
At the output of the laser, the beam is first passed through a triangular cavity with a 
roundtrip length of 20\,cm. The purpose of this mode cleaner cavity~\cite{Willke:98} is to suppress
spatial jitter in the beam, as this can lead to errors in the frequency locking of the beams to the 
gyroscope cavity, coupling directly into the rotation signal. The 
cavity's length is adjustable with a PZT-mounted end mirror. To lock the cavity length to the input beam, 
a 1\,MHz dither is applied to the PZT, and the photocurrent from the small leakage beam through the end 
mirror is mixed with the dither drive to provide a linear error signal. In essence, the mode cleaner converts 
beam jitter at its input into power fluctuations at the output. Completing the circle, the power fluctuations at the output of the mode cleaner are read out and fed back to an acousto-optic modulator and thereby suppressed. The length control loop for the mode cleaner has a bandwidth of a few\,kHz, which is sufficient to keep the RMS error well below its cavity linewidth.

After the mode cleaner, the beam passes through an electro-optic modulator (EOM) crystal that is 
resonant at 29.489\,MHz, which imparts phase modulation sidebands at this frequency for the PDH 
locking scheme. To minimize the effects of residual amplitude modulation 
(RAM)\,\cite{RAM:Keiko, wong1985servo, Hopper:2009a} 
from the crystal, the EOM is passively thermally isolated with insulating foam, and its metal housing 
is actively temperature stabilized above room temperature.

Following the EOM, the beam is split. The primary (CCW) beam is sent directly towards the cavity, while the 
secondary (CW) beam is upshifted by 100\,MHz using an AOM. To avoid strong beam jitter due to the 
first-order dependence of the Bragg scattering angle on the modulation frequency (which is varied minutely 
by the feedback), a double-pass scheme is used: on a first pass, the beam is upshifted by 50\,MHz; then, the 
beam is retro-reflected using a spherical mirror and passes through the AOM once more, acquiring a 
total roundtrip shift of 100\,MHz before being directed towards the cavity.

The geometry of the cavity dictates that the one beam's reflected path is co-spatial with the input 
path of the other. In order to separate the respective reflected beams, a polarization-isolation 
scheme is used: each beam passes through a Faraday isolator (FI) on its way to the cavity, and 
then half-wave plates are used to ensure that each reflected beam is rejected by the conjugate 
FI on return. Here, the two reflected fields are acquired on custom-built, low-noise RF 
photodetectors (RFPDs) to generate the locking error signals (these RFPDs are described 
in more detail below).

The photodetectors used for PDH locking of the cavity, as well as for the transmitted-beam beat 
readout, are custom-built versions of the design described by Grote~\cite{Grote:2007kl}. 
This RFPD topology is well suited for low-noise detection of narrowband RF signals in the 
presence of unwanted harmonics and with high DC power levels. In addition---unlike 
conventional resonant designs---it is not susceptible to photodiode bias modulation in the 
presence of large signals, which is particularly important in the case of the transmitted-beam readout,
since the heterodyne beat signal is not suppressed.

The primary loop, which locks the CCW beam to the cavity via feedback to the laser, can be considered 
to control the ``common mode'' of the system since actuation on the laser adjusts the frequency of 
both beams. As discussed in the noise analysis section below, this common degree of freedom must be 
controlled with high accuracy to prevent pollution of the differential mode that contains the rotation signal. 
Therefore, the primary loop is designed to be a high-performance system. After mixing and low-pass 
filtering using the PDH method, the derived error signal is passed through a custom-built multi-stage 
servo filter that allows for the very high 
low-frequency gains necessary to sufficiently suppress the common-mode noise. The output of the 
servo filter drives the NPRO PZT actuator to adjust the laser frequency. The primary loop achieves a 
unity-gain frequency (UGF) of $\sim\,15$\,kHz, limited by resonances in the NPRO PZT, with a 
resulting low-frequency loop gain of $>10^{12}$ below $\sim\,1$\,Hz.

The requirements on the secondary loop are not so stringent, since environmental noise is all but 
eliminated by the primary loop. In reality, the rotation signal can be read out from the secondary 
error signal while the loop is open. This mode of operation was tested, and in most cases a similar performance was achieved. However, this can lead to non-linearities at large signal levels 
and in general requires a more complicated calibration procedure. To lock this loop, the CW 
error signal is fed to another, lower-performance servo filter, whose output drives the external 
modulation input of the VCO driving the AOM. The actuation gain of the VCO is controllable using 
the external modulation input deviation range. The relative frequency noise of the VCO is 
proportional to this range, and therefore the minimal acceptable range is chosen based on 
the observed signal level. The secondary loop is chosen to have a UGF of $\sim\,20$\,kHz, 
and it provides loop gains of $>10^8$ below $\sim\,1$\,Hz.

\subsection{Realized performance and noise analysis}

\begin{figure}[htbp]
\begin{center}
\includegraphics[width=\columnwidth]{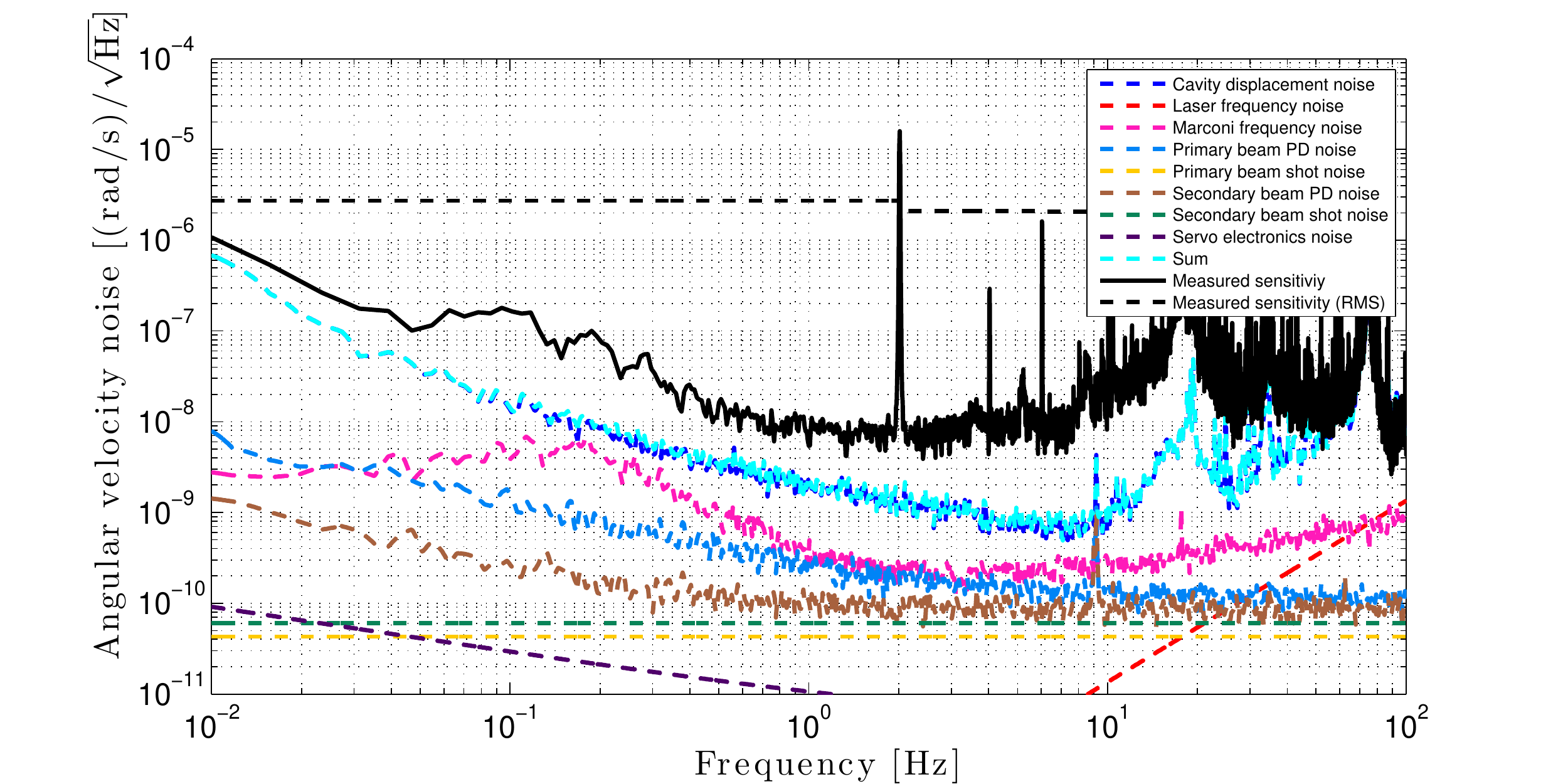}
\caption{Noise budget of the laser gyroscope. The measured sensitivity is plotted alongside the individual noise contributions. Mechanical noise is predicted to be the largest contributor, but the gyroscope exhibits excess noise above the expected level.}
\label{fig:nb}
\end{center}
\end{figure}

A noise budget for the passive laser gyroscope is in Fig.\,\ref{fig:nb}, showing the realized 
sensitivity of $10^{-8}$ (rad/s)/$\sqrt{\mbox{Hz}}$ above 500 mHz and increasing as 1/$f$ at lower frequencies. This section offers a detailed analysis of the various noise contributions.

When analyzing the performance of the instrument, it is helpful to first consider the ideal scenario. In that case, the primary loop has infinite gain at all frequencies, making the laser frequency perfectly follow the CCW mode of the cavity. Then, the secondary loop must only correct for differences between the CW and CCW modes. Were it not for the one-axial-mode shift between the beams' frequencies, any length fluctuations in the cavity would be completely common mode, and would therefore not couple to the rotation signal to first order\footnote{Since the so-called ``scale-factor'' (Eq.\,\ref{eq:rads2hz}) is dependent on the cavity length, there would still be a small, higher-order coupling.}. The fact that these two modes are separated in frequency leads to a residual length noise contribution with a common-mode rejection of $\nu_{\rm FSR} / \nu \approx 3.5 \times 10^{-7}$. 

Returning to the realistic limit of finite primary-loop gain, another source of noise emerges: since the primary loop cannot perfectly follow the CCW mode, any common-mode residuals must be corrected for by the secondary loop. This residual noise appears directly in the rotation signal, and it is the reason for the stringent primary-loop gain requirement. 

These are the only two noise sources somewhat unique to the gyroscope; as seen below, the remainder are more standard contributors.

\subsubsection{Mechanical noise}

As described briefly above, cavity length fluctuations driven by mechanical noise are largely common mode. However, the macroscopic frequency shift between the CW and CCW beams spoils the otherwise perfect cancellation. Allowing for some small fluctuation $\delta S$ in the cavity perimeter, we find that the supported eigenfrequencies in both directions are
\begin{eqnarray}
\nu_{\rm ccw} &=& n\frac{c}{S+\delta S} \\
\nu_{\rm cw} &=& (n+1)\frac{c}{S+\delta S},
\end{eqnarray}
where $n$ is an integer. From the raw beat signal, we will subtract the known offset of $c/S \approx 100$ MHz, giving a fluctuating frequency signal
\begin{equation} \label{eq:FSR_mod}
\Delta \nu_{\rm sig} = -\frac{c \,\delta S}{S^2 + S \,\delta S} \approx -c \frac{\delta S}{S^2},
\end{equation}
where the approximation $\delta S \ll S$ has been made. The linear term above is the first-order modulation of the FSR offset signal due to the cavity length fluctuation. Comparing it with the common-mode frequency shift from the same length fluctuation (i.e., $\Delta \nu_{\rm sig}^0 = -\nu \, \delta S / S$), we obtain the common-mode rejection ratio
\begin{equation}
{\rm CMRR} \equiv \frac{\Delta \nu_{\rm sig}}{\Delta \nu_{\rm sig}^0} \approx \frac{\nu_{\rm FSR}}{\nu} = \frac{\lambda}{S} \approx 3.5 \times 10^{-7}.
\end{equation}

This noise can be measured using using auxiliary channels and subtracted from the rotation signal, either online or in post-processing. To estimate the contribution, the actuation signal for the primary loop is monitored. Where that signal is dominated by the length fluctuations in the cavity (as opposed to, e.g., laser frequency noise), it is a faithful monitor of the common-mode noise. This signal is then multiplied by the CMRR, converted to rotation noise via Eq.\,\ref{eq:rads2hz}, and subtracted from the rotation signal.

In practice, below $\sim 1$ Hz, the free-running laser frequency noise is larger than the frequency equivalent of the cavity motion. Since laser frequency noise \emph{is} completely common mode, this can lead to an overestimation of the mechanical noise, and therefore to pollution of the rotation signal upon subtraction. To combat this, we have installed a laser frequency monitor by beating a pickoff of the gyroscope laser output with light that has been stabilized by locking to a quiet reference cavity. When the gyroscope is locked, this in-loop signal contains information only about the cavity motion, which allows for faithful subtraction.

\subsubsection{Residual common-mode noise}

The secondary loop simply acts to adjust its beam's frequency to match that mode's eigenfrequency. Therefore, any residual failure of the primary loop to lock the laser frequency to \emph{its} mode results in an injection of common-mode noise into the secondary loop (and hence directly into the rotation signal). Mathematically, the open-loop frequency error signal seen by the secondary loop is
\begin{equation} \label{eq:spill}
\nu_{\rm err} = \left( \frac{1}{1+G_{\rm p}} + {\rm CMRR} \right) \nu \frac{\delta S}{S} + \frac{4}{\lambda S} \vec A \cdot \vec \omega.
\end{equation}
The first term in the parentheses---where $G_{\rm p}$ is the primary loop gain---is the residual frequency error that remains after the action of the primary loop, while the second term is the differential mechanical noise described in the previous section. The last term is the Sagnac rotation signal.

In principle, this noise could be subtracted as well, but it is more effective to simply increase the primary loop gain until it is suppressed below the desired level. Since the target operational band is in the 10 mHz - 1 Hz range, it is easy to shape the primary servo filter to have an acceptable level of gain. For budgeting purposes, this noise contribution is calculated by measuring the in-loop primary error signal and referring it first to optical frequency (by dividing by the primary-loop optical gain in [V/Hz]) and then to rotation noise using Eq.\,\ref{eq:rads2hz}.

\subsubsection{Oscillator frequency noise}
\label{s:osc_freq}
As described in Sec.\,\ref{sec:overview}, the frequency stability of the RF VCOs used in the experiment is an
important factor in the ultimate sensitivity of the gyroscope\footnote{The stability of the fixed RF oscillator 
used to provide the PDH sidebands is comparatively unimportant, as the common-frequency, balanced-phase 
modulation/demodulation scheme gives first-order insensitivity to this oscillator noise.}. Both oscillators in this 
experiment are IFR/Marconi 2023A\,\cite{Marconi} RF sources, operated in external-input frequency 
modulation mode. For improved low-frequency stability, both sources are locked to a Stanford Research 
Systems FS725\,\cite{SRS_Rb} rubidium frequency standard via a 10-MHz reference signal. In the external-FM 
mode, the 2023A's frequency noise is proportional to the FM actuation range.

The VCO noise enters the rotation signal in different ways depending on the mode of operation. In the 
AOM actuation readout mode (see Sec.\,\ref{sec:overview}), the rotation signal is taken at the external FM 
input of VCO1. The secondary loop acts to cancel this noise at the optical error point, and therefore this 
signal contains directly the frequency noise of the VCO.

In the transmission beat note readout, on the other hand, the CW optical signal contains the much smaller, 
loop-suppressed contribution of the noise from VCO1. However, the noise from VCO2 is imposed directly 
on the PLL control signal readout in a similar fashion to above.

From one case to another, the required FM deviation range and the carrier frequency are only different by a 
factor of 2 (VCO1 is at $f_{\rm c1} = 50$\,MHz, due to the AOM double-pass setup, while VCO2 is at the 
full $f_{\rm c2} = 100$\,MHz.), and so the ultimate contribution to the rotation noise is roughly the same.

Since VCO1 is used as an actuator, there is little room to reduce the oscillator noise contribution in the 
first case. In the second case, however, VCO2 is only used as an actuator due to the PLL topology. Here, 
all that is necessary is a low-noise frequency discriminator, and so one is free to choose an alternate 
design with lower internal frequency noise. Due to the demanding relative frequency noise requirement 
of $10^{-11}/\sqrt{\mbox{Hz}}$ at and below 1\,Hz, several candidates (e.g., delay-line mixer 
frequency discriminator, LC detector, etc.) seem impractical~\cite{schilt:123116}. 
Finding a suitably stable frequency  discriminator is an important step to further improving the 
sensitivity of our design. 

\subsubsection{Electronic noise}

The electronic noise of each loop is measured independently and calibrated to units of rotation signal as appropriate. In order to measure this noise, the laser is blocked and the actuation signals from both loops are measured in this ``dark'' state. Both measured noise spectra are divided by their respective servo filter and optical gains to refer them to their inputs, and these form the effective sensing noise levels.

For the primary loop, this sensing noise sets a limit to the achievable common-mode noise suppression. In effect, it adds a fixed term $\delta \nu_{1}^{\rm sens}$ to Eq.\,\ref{eq:spill}, such that increasing $G_{\rm p}$ beyond a threshold level of
\begin{equation}
G_{\rm p}^{\rm SNL} \equiv \frac{\nu \,\delta S}{S \,\delta \nu_{1}^{\rm sens}} -1
\end{equation}
no longer reduces the contribution of the primary loop noise to the rotation signal. If $G_{\rm p} > G_{\rm p}^{\rm SNL}$, the primary loop is {\it sensing noise limited}. This threshold depends on the environmental noise present (through $\delta S$), but in practice $G_{\rm p}$ is high enough to meet this criterion at all times in the frequency range of interest.

Since the secondary loop reads the rotation signal out directly, sensing noise there constitutes an absolute limit for the rotational sensitivity in the most basic way, using Eq.\,\ref{eq:rads2hz}.

Using ultra-low-noise front-end electronics, and preferentially distributing gain upstream in the signal chain to reduce the effect of noisier later stages, the total contribution of electronic noise can be made negligible with respect to other sources. This is one motivation for using a high-finesse optical cavity, since a given voltage fluctuation in the front-end electronics corresponds to a smaller optical frequency fluctuation than with a lower-finesse cavity.

\subsubsection{Residual amplitude modulation (RAM)}
\label{sec:RAM}
As with any FM spectroscopy scheme, the PDH technique is susceptible to error from {\it residual amplitude modulation (RAM)} in the EOM used for control sideband generation. 
The EOM operates by applying an electric field across a crystal that exhibits the Pockels effect (i.e., it 
has a birefringence linearly proportional to the electric field applied). If the input beam polarization 
(and electrode axis) are exactly aligned with the appropriate crystal axis, the result is a pure phase modulation 
(PM) of the output beam. However, any slight misalignment can result in oscillatory rotation of the output beam 
polarization at the same frequency as the phase modulation. Upon interaction with any polarization-sensitive 
optics (e.g., polarizing beam splitters, birefringent mirrors, etc.), this polarization rotation is converted directly 
into amplitude modulation. As a result, the beams incident on the optical cavity have both PM and AM at the same frequency.

In the PDH scheme, a slight frequency offset from resonance converts PM into AM, which can be detected in the reflected beam by a photodetector, and the phase of this AM signal encodes the sign information about the offset. Therefore, any 
AM present in the input beam leads directly to an unwanted offset in the locking loop (i.e., the loop acts to cancel the 
RAM-induced offset by creating equal and opposite AM via frequency offset from the cavity eigenmode). 

A fixed AM level generates a static offset, which can be corrected for in principle by adding an electronic offset at 
the front end. A time-varying offset, however, is indistinguishable from a true signal, and in this way RAM directly 
produces noise in the rotation signal. 

The dominant driving source for low-frequency RAM fluctuations is temperature noise. For this reason, the EOM is 
temperature stabilized both passively and actively. For passive isolation, the EOM is covered by shape-fitting 
thermally insulating foam, which suppresses temperature-driven noise above the resultant thermal pole around 100\,mHz. To suppress noise below the thermal pole, the temperature of the EOM enclosure is actively stabilized using a custom 
controller.

\begin{figure}[htbp]
\begin{center}
\includegraphics[width=0.7\columnwidth]{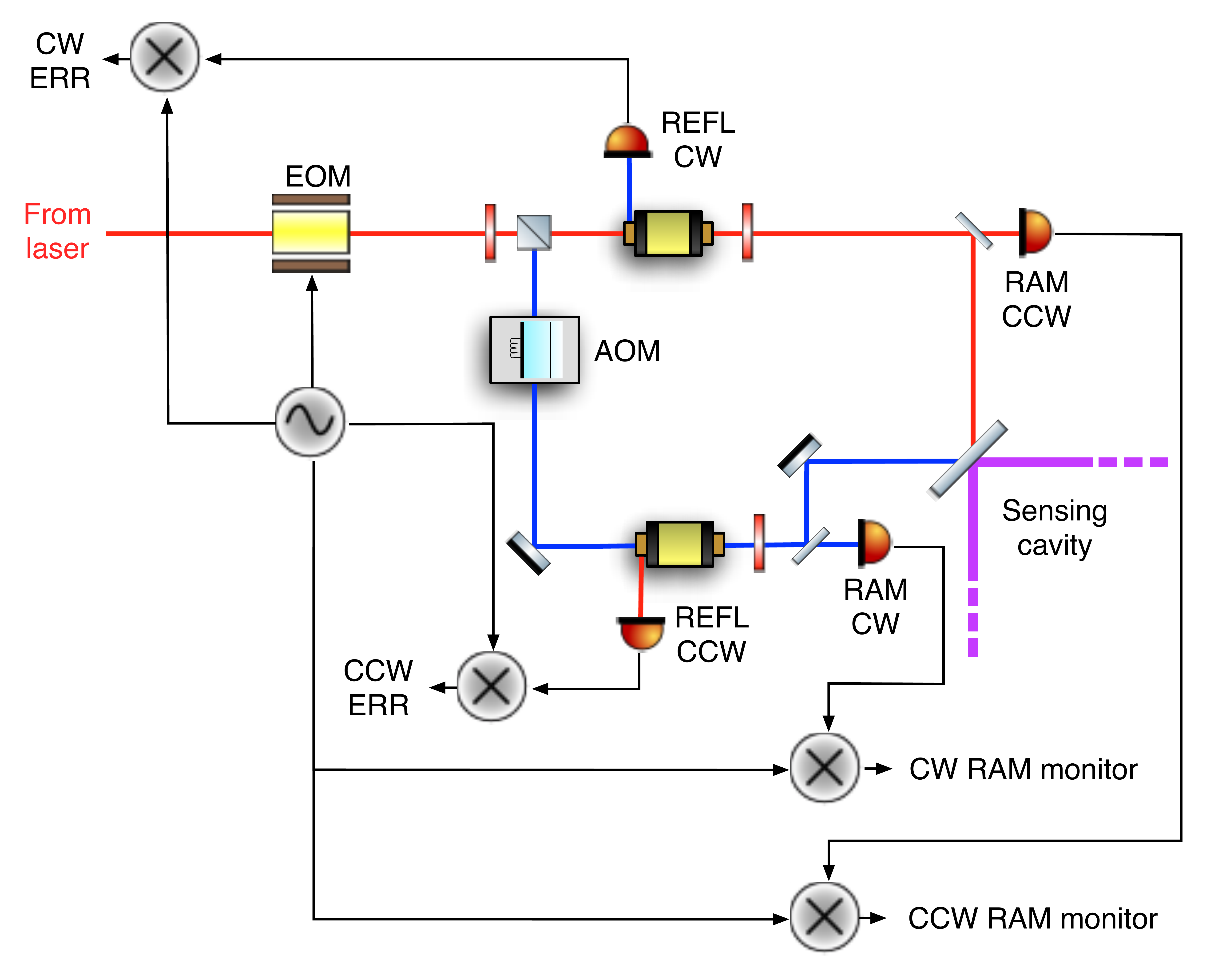}
\caption{Diagram of the out-of- RAM monitor setup.}
\label{fig:RAM_mon}
\end{center}
\end{figure}

This active/passive stabilization scheme is observed to suppress the RAM contribution considerably in broadband. 
However, even with this frontal suppression, the low-frequency rotation noise was often discovered to be coherent with 
RAM in at least one of the two beams. To further suppress this noise, an out-of-loop RAM witness photodetector 
was placed at a pickoff of each input path near the cavity injection (see Fig.\,\ref{fig:RAM_mon}). Since the beam 
sampled here has not yet interacted with the cavity, any AM present at the modulation frequency is RAM-induced. 
Each monitor photocurrent is mixed down with the PDH local oscillator signal---with the appropriate phase shift---in 
order to obtain the spurious RAM-induced component of the signal.

To chose the appropriate demodulation phase, the input polarization to the EOM is misaligned to introduce intentionally 
large RAM. Then, the demodulation phase is adjusted to maximize the DC output of the RAM monitors. Once this phase 
is set, the polarization is realigned to zero the RAM monitor signals, which are then amplified and digitally 
acquired for subsequent post-processing.

With the RAM monitor signals recorded along with the rotation signal, etc., Wiener filtering can subsequently be 
used to subtract any components of the rotation noise coherent with the RAM monitor signals.

\clearpage
\section{Possible future designs}
\label{s:future}

\begin{figure}[htbp]
\centering
	\includegraphics[width=\columnwidth]{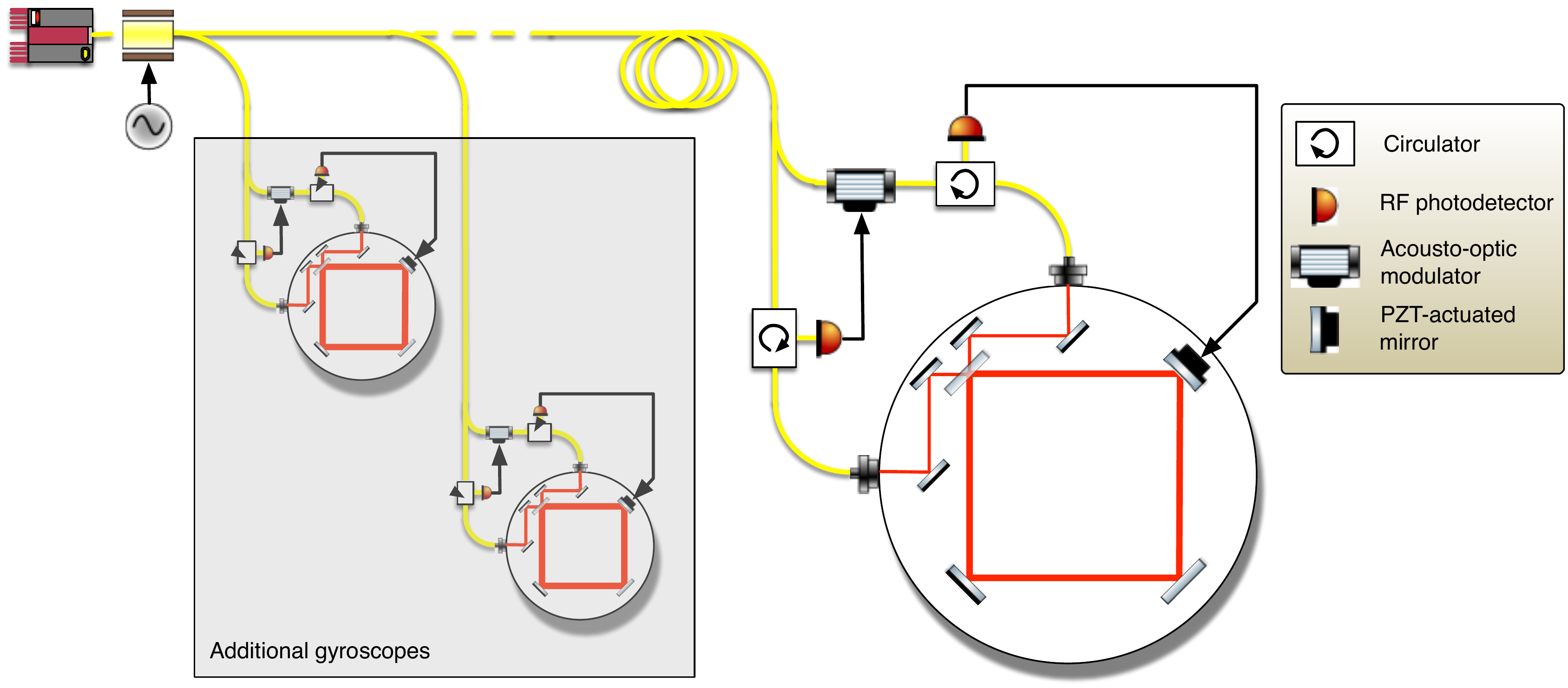}
	\caption{Proposed fiber-distributed gyroscope array scheme.}
	\label{fig:gyro_array}
\end{figure}

Given the information gleaned from this initial experiment, we have considered one 
possible future design. In doing so, we have paid particular attention to robustness, 
cost, and ability to be scaled up for use in demanding scientific environments.

In the foregoing analysis, it is clear that mechanical disturbances are a major source of noise. One obvious improvement, therefore, is to increase the mechanical stability of the ring cavity. This could be done by using a monolithic construction, where the cavity is of a single piece of a suitable low-expansion material. A superior solution involves changing the locking topology: by adding a length actuator to one cavity mirror, one can lock the cavity length to the laser, rather than the other way around. If the laser source is externally stabilized, it becomes a quiet reference against which the cavity length fluctuations are measured and strongly suppressed via feedback.

As an added advantage, using a quiet, fixed-frequency laser allows for multiple gyroscopes to be illuminated by the same source. By taking advantage of this fact, one could create a relatively inexpensive array of gyroscopes. Each individual unit would consist of its own cavity, two photodetectors, and an acousto-optic modulator to internally generate its secondary beam. The central laser would be distributed to each gyroscope using a robust system of optical fibers, and each unit would be pre-aligned internally, allowing for easy removal and reconnection of the light source for unit relocation.

Since this scheme would strongly suppress the mechanical noise that dominates in our current system, the ring cavity could also be reduced in size by a factor of two or more while maintaining improved sensitivity. As per Eq.\,\ref{eq:rads2hz}, the rotation sensitivity for a given readout frequency noise level scales linearly with the cavity length. Therefore, for example, the area may be scaled down by a factor of four, and the resulting projected noise floor would be only a factor of two higher than the magenta trace in Fig.\,\ref{fig:nb} (since this is the next dominant noise term after mechanical noise). In addition to making the design more compact, this may also facilitate making the monolithic cavity structure highly mechanically stable.

\clearpage
\section{Conclusion}
\label{s:conclusion}

We have described a novel laser gyroscope design employing a passive, free-space 
optical cavity. By separating the counter-propagating sensing optical fields macroscopically 
in frequency, this design is free from the lock-in 
effects observed in previous systems (active and passive). We have 
demonstrated a rotation sensitivity of $10^{-7} \mbox{ radians}/\sqrt{\mbox{Hz}}$ 
below 1\,Hz. Our instrument noise floor is compared with those of other contemporary rotation sensors in Fig.\,\ref{fig:comp_plot}. A convenient feature of this design is the ability to use a 
commercial laser to illuminate the system, rather than building a 
custom laser resonator.


We believe our design lends itself to integration in complex systems 
requiring accurate rotation sensing. In particular, the sensing 
cavity need not be an ad-hoc mirror-and-mount structure; we envision 
that a future version could benefit from a monolithic design, 
which could make the system more compact as well as more immune
to environmental disturbances.

\section{Acknowledgements}
Thanks to grants numbers, Brian Lantz for early suggestions,
Shaoul Ezekiel for comments on the early setup, members of the
LSC SWG. Thanks as well to the Caltech SURF program, and in particular to Michelle Stephens, Jenna Walrath and Zoe Masters for their contributions as summer fellows. Finally, we are grateful for help and advice from Denis Martynov, Tara Chalermsomsak, Frank Seifert, Jenne Driggers, David Yeaton-Massey, and Steve Vass.

\section*{References}
\bibliographystyle{iopart-num}
\bibliography{gyro_bib}

\end{document}